\pdfoutput=1 

\documentclass[12pt,a4paper]{article}

\usepackage{ifthen} 
\newboolean{pdflatex}
\setboolean{pdflatex}{true} 

\newboolean{articletitles}
\setboolean{articletitles}{true} 

\newboolean{uprightparticles}
\setboolean{uprightparticles}{false} 

\def\paperauthors{LHCb collaboration} 
\def\paperasciititle{Measurement of charged-hadron production in Z-tagged jets} 
\def\papertitle{Measurement of charged hadron production in \Z-tagged jets in proton-proton collisions at $\sqs=8\tev$} 
\def\paperkeywords{{High Energy Physics}, {LHCb}} 
\def\papercopyright{\the\year\ CERN for the benefit of the LHCb collaboration} 
\def\paperlicence{CC-BY-4.0 licence}
\def\paperlicenceurl{https://creativecommons.org/licenses/by/4.0/}

\usepackage[top=1in, bottom=1.25in, left=1in, right=1in]{geometry}

\usepackage{dcolumn,booktabs}
\newcolumntype{d}[1]{D{.}{.}{#1}}

\usepackage{microtype}
\usepackage{lineno}  
\usepackage{xspace} 
\usepackage{caption} 

\usepackage{graphicx}  
\usepackage{color}
\usepackage{colortbl}
\graphicspath{{./figs/}} 
\DeclareGraphicsExtensions{.pdf,.PDF,png,.PNG}

\usepackage{amsmath} 
\usepackage{amssymb}
\usepackage{amsfonts}
\usepackage{upgreek} 

\newcommand*\patchAmsMathEnvironmentForLineno[1]{%
\expandafter\let\csname old#1\expandafter\endcsname\csname #1\endcsname
\expandafter\let\csname oldend#1\expandafter\endcsname\csname
end#1\endcsname
 \renewenvironment{#1}%
   {\linenomath\csname old#1\endcsname}%
   {\csname oldend#1\endcsname\endlinenomath}%
}
\newcommand*\patchBothAmsMathEnvironmentsForLineno[1]{%
  \patchAmsMathEnvironmentForLineno{#1}%
  \patchAmsMathEnvironmentForLineno{#1*}%
}
\AtBeginDocument{%
\patchBothAmsMathEnvironmentsForLineno{equation}%
\patchBothAmsMathEnvironmentsForLineno{align}%
\patchBothAmsMathEnvironmentsForLineno{flalign}%
\patchBothAmsMathEnvironmentsForLineno{alignat}%
\patchBothAmsMathEnvironmentsForLineno{gather}%
\patchBothAmsMathEnvironmentsForLineno{multline}%
\patchBothAmsMathEnvironmentsForLineno{eqnarray}%
}

\usepackage{hyperxmp}

\usepackage[pdftex,
            pdfauthor={\paperauthors},
            pdftitle={\paperasciititle},
            pdfkeywords={\paperkeywords},
            pdfcopyright={Copyright (C) \papercopyright},
            pdflicenseurl={\paperlicenceurl}]{hyperref}

\usepackage[colorinlistoftodos,textsize=scriptsize]{todonotes}

\usepackage[all]{hypcap} 

\usepackage{xspace} 
\usepackage{upgreek}

\def\lhcb   {\mbox{LHCb}\xspace}
\def\atlas  {\mbox{ATLAS}\xspace}

\def\MagUp {\mbox{\em Mag\kern -0.05em Up}\xspace}

\ifthenelse{\boolean{uprightparticles}}%
{

 \def\PDelta      {\ensuremath{\Delta}\xspace}                 
 \def\PXi         {\ensuremath{\Xi}\xspace}                 
 \def\PLambda     {\ensuremath{\Lambda}\xspace}                 
 \def\PSigma      {\ensuremath{\Sigma}\xspace}                 
 \def\POmega      {\ensuremath{\Omega}\xspace}                 
 \def\PUpsilon    {\ensuremath{\Upsilon}\xspace}

 \def\PB      {\ensuremath{\mathrm{B}}\xspace}                 
                  
 \def\PD      {\ensuremath{\mathrm{D}}\xspace}

 \def\PK      {\ensuremath{\mathrm{K}}\xspace}

 \def\PZ      {\ensuremath{\mathrm{Z}}\xspace}

 \def\Pi      {\ensuremath{\mathrm{i}}\xspace}

 \def\thebaroffset{0.0em}
}
{

 \mathchardef\PDelta="7101
 \mathchardef\PXi="7104
 \mathchardef\PLambda="7103
 \mathchardef\PSigma="7106
 \mathchardef\POmega="710A
 \mathchardef\PUpsilon="7107
                  
 \def\PB      {\ensuremath{B}\xspace}                 
                  
 \def\PD      {\ensuremath{D}\xspace}

 \def\PK      {\ensuremath{K}\xspace}

 \def\PZ      {\ensuremath{Z}\xspace}

 \def\Pi      {\ensuremath{i}\xspace}

 \def\thebaroffset{0.18em}
}
\newcommand{\offsetoverline}[2][\thebaroffset]{\kern #1\overline{\kern -#1 #2}}%

\makeatletter
\ifcase \@ptsize \relax
  \newcommand{\miniscule}{\@setfontsize\miniscule{4}{5}}
\or
  \newcommand{\miniscule}{\@setfontsize\miniscule{5}{6}}
\or
  \newcommand{\miniscule}{\@setfontsize\miniscule{5}{6}}
\fi
\makeatother

\DeclareRobustCommand{\optbar}[1]{\shortstack{{\miniscule (\rule[.5ex]{1.25em}{.18mm})}
  \\ [-.7ex] $#1$}}

\def\Z      {{\ensuremath{\PZ}}\xspace}

\def\KorKbar {\kern \thebaroffset\optbar{\kern -\thebaroffset \PK}{}\xspace}

\def\DorDbar {\kern \thebaroffset\optbar{\kern -\thebaroffset \PD}\xspace}

\def\BorBbar {\kern \thebaroffset\optbar{\kern -\thebaroffset \PB}\xspace}

\def\Y#1S{\ensuremath{\PUpsilon{(#1S)}}\xspace}

\def\LorLbar     {\kern \thebaroffset\optbar{\kern -\thebaroffset \PLambda}\xspace}

\def\AT#1     {\ensuremath{A_{\mathrm{T}}^{#1}}\xspace}           

\def\C#1      {\ensuremath{\mathcal{C}_{#1}}\xspace}                       
\def\Cp#1     {\ensuremath{\mathcal{C}_{#1}^{'}}\xspace}                    
\def\Ceff#1   {\ensuremath{\mathcal{C}_{#1}^{\mathrm{(eff)}}}\xspace}        
\def\Cpeff#1  {\ensuremath{\mathcal{C}_{#1}^{'\mathrm{(eff)}}}\xspace}       
\def\Ope#1    {\ensuremath{\mathcal{O}_{#1}}\xspace}                       
\def\Opep#1   {\ensuremath{\mathcal{O}_{#1}^{'}}\xspace}                    

\newcommand{\aunit}[1]{\ensuremath{\text{\,#1}}}       

\newcommand{\tev}{\aunit{Te\kern -0.1em V}\xspace}
\newcommand{\gev}{\aunit{Ge\kern -0.1em V}\xspace}
\newcommand{\mev}{\aunit{Me\kern -0.1em V}\xspace}
\newcommand{\kev}{\aunit{ke\kern -0.1em V}\xspace}
\newcommand{\ev}{\aunit{e\kern -0.1em V}\xspace}
\newcommand{\mevc}{\ensuremath{\aunit{Me\kern -0.1em V\!/}c}\xspace}
\newcommand{\gevc}{\ensuremath{\aunit{Ge\kern -0.1em V\!/}c}\xspace}
\newcommand{\mevcc}{\ensuremath{\aunit{Me\kern -0.1em V\!/}c^2}\xspace}
\newcommand{\gevcc}{\ensuremath{\aunit{Ge\kern -0.1em V\!/}c^2}\xspace}

\def\fb   {\ensuremath{\aunit{fb}}\xspace}
\def\invfb   {\ensuremath{\fb^{-1}}\xspace}

\def\gsim{{~\raise.15em\hbox{$>$}\kern-.85em
          \lower.35em\hbox{$\sim$}~}\xspace}
\def\lsim{{~\raise.15em\hbox{$<$}\kern-.85em
          \lower.35em\hbox{$\sim$}~}\xspace}

\def\sqs   {\ensuremath{\protect\sqrt{s}}\xspace}

\def\pt         {\ensuremath{p_{\mathrm{T}}}\xspace}

\def\evtgen     {\mbox{\textsc{EvtGen}}\xspace}

\def\geant      {\mbox{\textsc{Geant4}}\xspace}

\def\photos     {\mbox{\textsc{Photos}}\xspace}

\def\pythia     {\mbox{\textsc{Pythia}}\xspace}

\xspace

\def\tell1  {TELL1\xspace}
\def\ukl1   {UKL1\xspace}

\newcommand{\eg}{\mbox{\itshape e.g.}\xspace}

\newcommand{\zjet}{\mbox{\Z{}+jet}\xspace}
\newcommand{\jt}{\mbox{$j_{\textrm{T}}$}\xspace}

\usepackage{cite} 
\usepackage{mciteplus}

\begin{document}

\renewcommand{\thefootnote}{\fnsymbol{footnote}}
\setcounter{footnote}{1}



\begin{titlepage}
\pagenumbering{roman}

\vspace*{-1.5cm}
\centerline{\large EUROPEAN ORGANIZATION FOR NUCLEAR RESEARCH (CERN)}
\vspace*{1.5cm}
\noindent
\begin{tabular*}{\linewidth}{lc@{\extracolsep{\fill}}r@{\extracolsep{0pt}}}
\ifthenelse{\boolean{pdflatex}}
{\vspace*{-1.5cm}\mbox{\!\!\!\includegraphics[width=.14\textwidth]{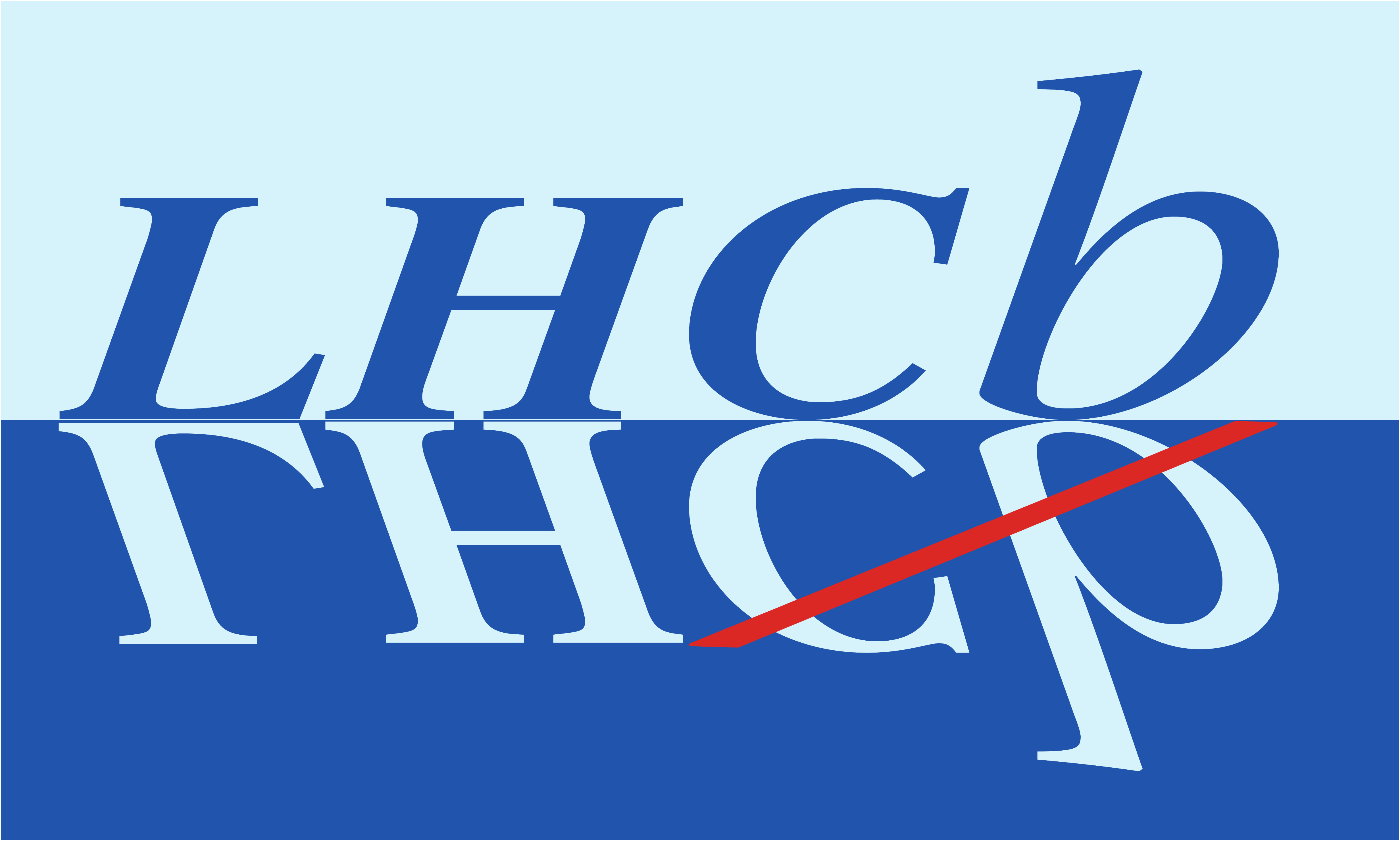}} & &}%
{\vspace*{-1.2cm}\mbox{\!\!\!\includegraphics[width=.12\textwidth]{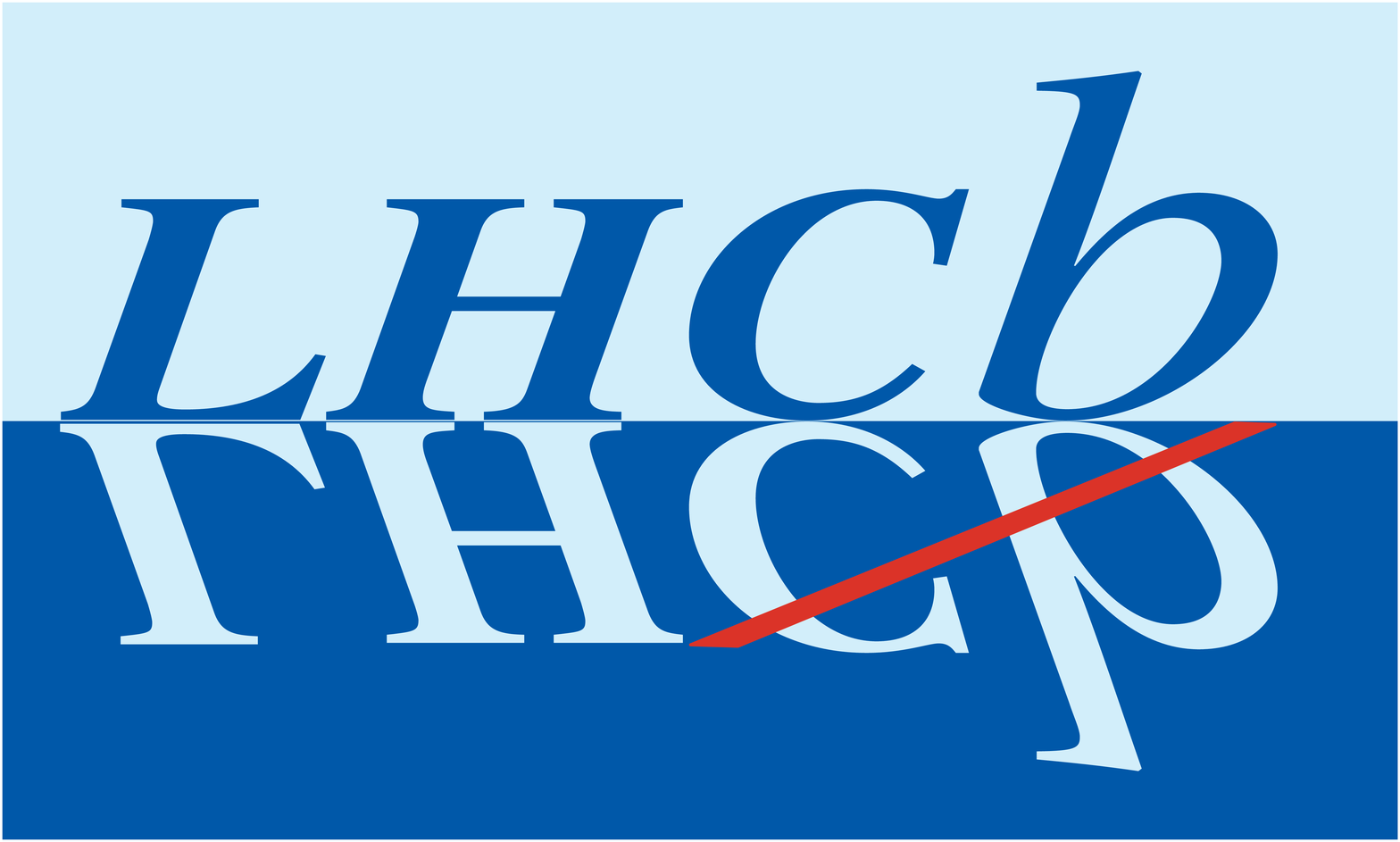}} & &}%
\\
 & & CERN-EP-2019-065 \\  
 & & LHCb-PAPER-2019-012 \\  
 & & \today \\ 
 & & \\
\end{tabular*}

\vspace*{2.5cm}

{\normalfont\bfseries\boldmath\huge
\begin{center}
  \papertitle 
\end{center}
}

\vspace*{1.5cm}

\begin{center}

\paperauthors\footnote{Authors are listed at the end of this paper.}
\end{center}

\vspace{\fill}

\begin{abstract}
  \noindent The production of charged hadrons within jets recoiling against a \Z boson is measured in proton-proton collision data at $\sqs=8\tev$ recorded by the \lhcb experiment. The charged-hadron structure of the jet is studied longitudinally and transverse to the jet axis for jets with transverse momentum $\pt>20\gev$ and in the pseudorapidity range $2.5<\eta<4$. These are the first measurements of jet hadronization at these forward rapidities and also the first where the jet is produced in association with a \Z boson. In contrast to previous hadronization measurements at the Large Hadron Collider, which are dominated by gluon jets, these measurements probe predominantly light-quark jets which are found to be more longitudinally and transversely collimated with respect to the jet axis when compared to the previous gluon dominated measurements. Therefore, these results provide valuable information on differences between quarks and gluons regarding nonperturbative hadronization dynamics.

\end{abstract}

\vspace*{2.0cm}

\begin{center}
 Published in Phys. Rev. Lett. 123 (2019) 232001
\end{center}

\vspace{\fill}

{\footnotesize 
\centerline{\copyright~\papercopyright. \href{\paperlicenceurl}{\paperlicence}.}}
\vspace*{2mm}

\end{titlepage}


\newpage
\setcounter{page}{2}
\mbox{~}

\cleardoublepage


\renewcommand{\thefootnote}{\arabic{footnote}}
\setcounter{footnote}{0}


\pagestyle{plain} 
\setcounter{page}{1}
\pagenumbering{arabic}


Quantum chromodynamics (QCD), the theory of the strong interaction, is unique amongst the fundamental forces due to the nonperturbative processes that confine quarks and gluons, collectively referred to as partons, within bound-state hadrons. The parton structure of protons has been the focus of intense research efforts; however, the understanding of how hadrons arise from scattered partons is limited in comparison. Perturbative QCD calculations utilize fragmentation functions to determine cross-sections of hadron production from scattered partons. Fragmentation functions describe the probability for a particular parton to transform into a particular hadron~\cite{Collins:1981uk,Collins:1981uw,Collins:2011zzd}. Several global fits to experimental data have provided parameterized fragmentation functions (see \eg Ref.~\cite{Metz:2016swz} and references therein). However, there is a significant lack of understanding in the mechanisms through which hadrons are formed in the nonperturbative hadronization process and therefore additional data are required.

Fragmentation function studies have been performed using inclusive hadron production at $e^+e^-$ colliders, which benefit from a simpler environment free of initial-state gluon radiation~\cite{Buskulic:1994ft,Akers:1994ez,OPAL:1995ab,Abreu:1998vq,Abe:1998zs,Abreu:1999af,Lees:2013rqd,Leitgab:2013qh,Seidl:2019jei}. Semi-inclusive deep-inelastic-scattering measurements have also been used to constrain fragmentation functions at smaller values of $Q^2$, the momentum transfer~\cite{Airapetian:2012ki,Aghasyan:2017ctw}. Additionally, inclusive hadron production measurements have been used to study fragmentation functions in the more complex environment, relative to interactions involving leptons, of proton-proton ($pp$) collisions~\cite{Adams:2006nd,Arsene:2007jd,Adare:2014wht}. However, such measurements are limited by the lack of an explicit way to relate the scattered parton to the final-state hadron. Measuring fragmentation functions with respect to high transverse momentum (\pt) jets offers a unique opportunity to study hadron production relative to an object that is correlated to the scattered parton. For example, the transverse profile, in addition to the longitudinal dynamics of hadrons within jets, can be used to study fragmentation functions in the longitudinal and transverse directions with respect to the jet axis. Such multidimensional measurements that go beyond inclusive hadrons, or those that consider correlations between particles, have the potential to answer unique questions within QCD related to universality, factorization, and the importance of color-charge flow~\cite{Collins:2002kn,Rogers:2010dm}.

This Letter reports a study of charged hadrons produced in jets recoiling against a \Z boson, also referred to as \Z-tagged jets, in the forward region of $pp$ collisions.\footnote{Throughout this Letter the notation \Z includes both the $Z^0$ and virtual $\gamma^*$ contributions.} The longitudinal momentum fraction, $z$, the momentum transverse to the jet axis, \jt, and the radial distribution, $r$, of charged hadrons are measured with respect to the jet axis in the laboratory frame, defined as
\begin{equation}
    z\equiv\frac{{\bf{p}}_{\textrm{jet}}\cdot{\bf{p}}_{\textrm{hadron}}}{|{\bf{p}}_{\textrm{jet}}|^2},
\end{equation}
\begin{equation}\label{eq:jt}
    \jt\equiv\frac{|{\bf{p}}_{\textrm{jet}}\times{\bf{p}}_{\textrm{hadron}}|}{|{\bf{p}}_{\textrm{jet}}|},
\end{equation}
and
\begin{equation}\label{eq:r}
r\equiv\sqrt{(\phi_{\textrm{jet}}-\phi_{\textrm{hadron}})^2+(y_{\textrm{jet}}-y_{\textrm{hadron}})^2}.
\end{equation}
Here, ${\bf{p}}$ is the 3-momentum vector, $\phi$ is the azimuthal angle, and $y$ is the rapidity. The data sample is selected from an integrated luminosity of approximately 2\invfb collected at a center-of-mass energy $\sqs=8\tev$ with the \lhcb detector in 2012. Events with only one reconstructed primary vertex are analyzed to better identify signatures of a hard two-to-two partonic scattering. Jets are clustered with the anti-$k_T$ algorithm~\cite{Cacciari:2008gp} using a distance parameter $R=0.5$ and are measured differentially in \pt for $\pt>20\gev$, and in the pseudorapidity range $2.5<\eta<4$.\footnote{In this Letter, natural units ($c=\hbar=1$) are used.} Charged hadrons within the jet are required to have $\pt>0.25\gev$, momentum $p>4\gev$, and to lie within the jet cone such that $\Delta R<0.5$, where $\Delta R \equiv \sqrt{(\phi_{\textrm{jet}}-\phi_{\textrm{hadron}})^2+(\eta_{\textrm{jet}}-\eta_{\textrm{hadron}})^2}$. The distributions are unfolded to account for the detector response and to facilitate comparisons with theoretical and numerical predictions. This is the first measurement of charged hadrons within jets produced in association with a \Z boson, as well as the first measurement of charged hadrons in jets at these forward pseudorapidities. The \zjet process is primarily sensitive to light quark jets, as demonstrated by PYTHIA in this kinematic range~\cite{Sjostrand:2006za,Sjostrand:2007gs}. Thus, these data provide new and complementary information to previous jet substructure measurements in the inclusive jet channel at midrapidity in hadronic collisions, which are sensitive to primarily gluon jets~\cite{Aad:2011sc,Chatrchyan:2014ava,ALICE:2014dla,Aaboud:2017tke,Aaboud:2017qwh,Acharya:2018eat}. Recent results at midrapidity in the isolated photon-jet channel can also probe fragmentation differences when a photon, rather than a massive vector boson, is measured opposite the jet~\cite{Aaboud:2019oac}.

The \lhcb detector is a single-arm forward spectrometer covering the pseudorapidity range $2<\eta<5$, described in detail in Refs.~\cite{Alves:2008zz,LHCb-DP-2014-002}. Simulations are used to evaluate the detector performance with regard to the jet reconstruction, track-in-jet reconstruction, and to validate the analysis methods. The simulated $pp\rightarrow Z+\textrm{jet}+X$ events are generated using \pythia8~\cite{Sjostrand:2007gs} with a specific \lhcb configuration~\cite{LHCb-PROC-2010-056}. Decays of hadronic particles are described by \evtgen~\cite{Lange:2001uf}, and final-state radiation in the simulation is generated using \photos~\cite{Golonka:2005pn}. Finally, the \geant toolkit~\cite{Agostinelli:2002hh, *Allison:2006ve} is used to simulate the interactions of the particles with the detector, as described in Ref.~\cite{LHCb-PROC-2011-006}.

This analysis uses the same data set as that used for the \zjet cross section measurement, where events are selected and \Z bosons are measured via their dimuon decay as described in Ref.~\cite{LHCb-PAPER-2016-011}. Candidate events are required to pass a trigger~\cite{LHCb-DP-2012-004} which selects muons with $\pt>10\gev$. Only events that contain two high-\pt muons are retained. The muons are required to satisfy track-reconstruction and muon-identification  criteria, as in Refs.~\cite{LHCb-PAPER-2015-049,LHCb-PAPER-2016-011}, and are also required to fall within the fiducial region of $2<\eta<4.5$, where the detector performance is well understood. Finally, the dimuon system must have an invariant mass, $M_{\mu\mu}$, within the range $60<M_{\mu\mu}<120\gev$. 

Jet reconstruction is performed using a particle flow algorithm~\cite{LHCb-PAPER-2013-058}, where the charged and neutral particles are clustered using the anti-k$_T$ algorithm as implemented in Ref.~\cite{Cacciari:2011ma}. Reconstructed jets with $\pt>15\gev$ that lie within $2.5<\eta<4$ are analyzed. The $15<\pt<20\gev$ region is included to avoid inefficiencies at the lower \pt limit of the measurement in the unfolding procedure. The pseudorapidity requirement ensures that the full jet cone lies within the fiducial area of the \lhcb detector. Selection requirements are placed on the jets to reduce the rate of jets not associated with the partonic process producing the \Z boson, which is already suppressed by the requirement of a single reconstructed primary vertex in the event. Additionally, the decay muons from the \Z boson must not be contained within the jet cone. Only jets that are on the azimuthal away-side of the \Z boson, defined by $\Delta\phi_{\Z-\textrm{jet}}\equiv|\phi_{\Z}-\phi_{\textrm{jet}}| > 7\pi/8$, are analyzed. The jet energy calibrations are the same as those used in Ref.~\cite{LHCb-PAPER-2016-011}. Charged hadrons within the jet are identified by the particle flow algorithm utilizing the particle-identification systems and several track-quality criteria~\cite{LHCb-DP-2014-002}. The charged hadrons must also satisfy $\Delta R<0.5$, which ensures that the corresponding tracks fall within the tracking acceptance.

The methods used to determine the charged-hadron fragmentation distributions are as described in Refs.~\cite{Aad:2011sc,LHCb-PAPER-2016-064,Aaboud:2017tke}. The fragmentation distributions are corrected for tracking inefficiencies and two-dimensionally unfolded for resolution effects, which primarily occur due to the jet energy resolution. The unfolded fragmentation distributions are then normalized by the total number of \Z{}+jet events in a given jet \pt bin, which is determined separately from the hadron-in-jet unfolding procedure. A test of the method described here and below, performed with the reconstructed simulation samples, confirmed that the generated distributions were reproduced for all observables studied, within the statistical uncertainties of the simulated sample. In this analysis, the \Z-boson \pt is integrated to provide the statistical precision to measure the fragmentation as a function of jet \pt. The integral of the fragmentation distributions then corresponds to the mean multiplicity of charged hadrons within the jet.

The number of \zjet pairs in each jet \pt bin is corrected to account for reconstruction and selection inefficiencies, and is determined independently from and normalizes the fragmentation distributions. The muon detection efficiencies are determined in data using the technique employed in the inclusive weak boson cross-section measurements of \lhcb~\cite{LHCb-PAPER-2015-001,LHCb-PAPER-2015-049}. The jet reconstruction efficiency is evaluated from simulation, and is greater than 90\% for jets with $\pt>20\gev$. A correction is also applied to account for differences between the number of events produced and measured in a given \pt bin due to the jet \pt resolution. This correction is determined from simulation, and is less than 10\%. The method described above is cross checked by comparing the results to a full Bayesian unfolding~\cite{DAgostini:1994fjx} as implemented in Ref.~\cite{Adye:2011gm}. The two methods agree to within 1\%.

Simulation is used to determine the tracking efficiency and to account for effects from misreconstructed tracks that are incorrectly measured inside or outside of the jet cone. The efficiency is evaluated as a function of momentum and pseudorapidity and applied on a per-track basis. The efficiency decreases for $p>150\gev$ due to a requirement on the uncertainty of the track bending radius which is part of the particle flow algorithm and has a larger effect at high momentum. The efficiencies were found to be independent of the observables $z$, $\jt$ and $r$ in the simulated sample. To validate the efficiency corrections procedure, the simulation sample is split in half and the efficiencies are determined with one half and applied to the other. Good recovery of the generated charged hadron distributions in $p$ and $\eta$ is observed. Within the statistical precision of the sample the tracking efficiency does not depend on the jet \pt.

The effects of bin migration in jet \pt and in the fragmentation observables on the fragmentation distributions, primarily due to the jet energy and momentum resolutions, are corrected using the two-dimensional Bayesian unfolding method. Response matrices are constructed for each fragmentation observable using simulated samples that study the correlations between the generated and reconstructed yields in bins of $[z,p_{\textrm{T}}^{\textrm{jet}}]$, $[\jt,p_{\textrm{T}}^{\textrm{jet}}]$, and $[r,p_{\textrm{T}}^{\textrm{jet}}]$. Typically the bin migration is less than 5\%; however, it can be larger for more extreme values of the fragmentation variables, for example at large $z$. The number of iterations in the Bayesian unfolding procedure is selected to be the minimum number for which the relative change in the fragmentation functions at $z\approx0.05$ is smaller than 0.2\% per additional iteration in all of the jet \pt bins. Based on this criterion, the unfolding is iterated seven times for each observable.

Systematic uncertainties that arise from the uncertainties on the various efficiencies are assigned to the number of \zjet pairs measured in each jet \pt bin. The uncertainty in the muon reconstruction efficiency is negligible. Systematic uncertainties on the jet reconstruction are evaluated as in  Ref.~\cite{LHCb-PAPER-2016-011} by comparing the jet reconstruction quality requirements in simulation and data. Similarly to the muon efficiencies, the precision with which the uncertainty of the jet reconstruction corrections are determined, due to the limited simulation sample size, is also evaluated; however, this is found to be negligible compared to the jet reconstruction quality requirement uncertainty of 1.9\%. The normalization is not corrected for \zjet background events, and thus a systematic uncertainty of 1.7\% is assigned for the impurity of both \Z bosons and jets in the measurement, as determined in Refs.~\cite{LHCb-PAPER-2015-049,LHCb-PAPER-2016-011}. Effects from pile up are also studied and found to be negligible. The total normalization uncertainty of 2.7\% is determined by adding these components in quadrature.

The jet-energy scale and resolution are also considered as sources of systematic uncertainty. The jet-energy scale and its uncertainty have been studied in previous measurements of the \zjet cross-section~\cite{LHCb-PAPER-2013-058,LHCb-PAPER-2016-011}. To estimate these effects, the scale is varied by one standard deviation of its uncertainty. New unfolding matrices are constructed with this modification, and the difference in the fragmentation distributions determined with the modified and nominal response matrices is taken as a systematic uncertainty. Similarly, the systematic uncertainty due to the jet-energy resolution is evaluated by smearing each  component of the jet momentum by an additional term corresponding to the uncertainty on the jet resolution and constructing new response matrices. The difference between the nominal and smeared unfolded charged hadron-in-jet distributions is taken as the systematic uncertainty on the jet-energy resolution.

The unfolding method is validated with two different tests. The first test is performed by splitting the simulated sample in two and using one half to generate the response matrices with which the other half is unfolded. Recovery of the generator-level fragmentation distributions is observed and average deviations from perfect agreement are 2\%, which is assigned as an uncertainty related to the unfolding procedure. A second test is performed by splitting the simulated sample in half by \Z-boson \pt, and performing a similar test to the previous one to check for any uncertainty associated with the assumed prior. The results again deviate from perfect agreement by about 2\%, confirming that a 2\% systematic uncertainty on the unfolding procedure is appropriate.

The track selection requirements, tracking efficiency, and charged-hadron identification are also studied as sources of systematic uncertainty. The track selection uncertainty is assigned by requiring a tight fake-track removal criterion and repeating the analysis. The differences in the final fragmentation distributions with and without this requirement are taken as systematic uncertainties. The track selection uncertainty is typically less than 5\%; however, it reaches a maximum of approximately 8\% at some values of $z$, $\jt$, and $r$ in the highest jet \pt bin studied. The systematic uncertainty on the tracking efficiency is determined by smoothing the two-dimensional efficiency and repeating the analysis, which accounts for the statistical precision with which the efficiency is determined. The resulting distributions are compared to the nominal distributions and the differences are taken as uncertainties on the tracking efficiency; these are generally less than 3\% but rise up to 10\% in some bins. Uncertainties associated to misidentifying charged hadrons are also considered by comparing the nominal fragmentation functions to those obtained when hadron-to-lepton (and \textit{vice versa}) misidentification probabilities are considered. These uncertainties are less than 5\%, except at large $z$ where the charged pion-to-electron misidentification probability becomes larger~\cite{LHCb-DP-2014-002}.

%
%
\begin{figure}[tb]
    \centering
    \includegraphics[width=0.6\linewidth]{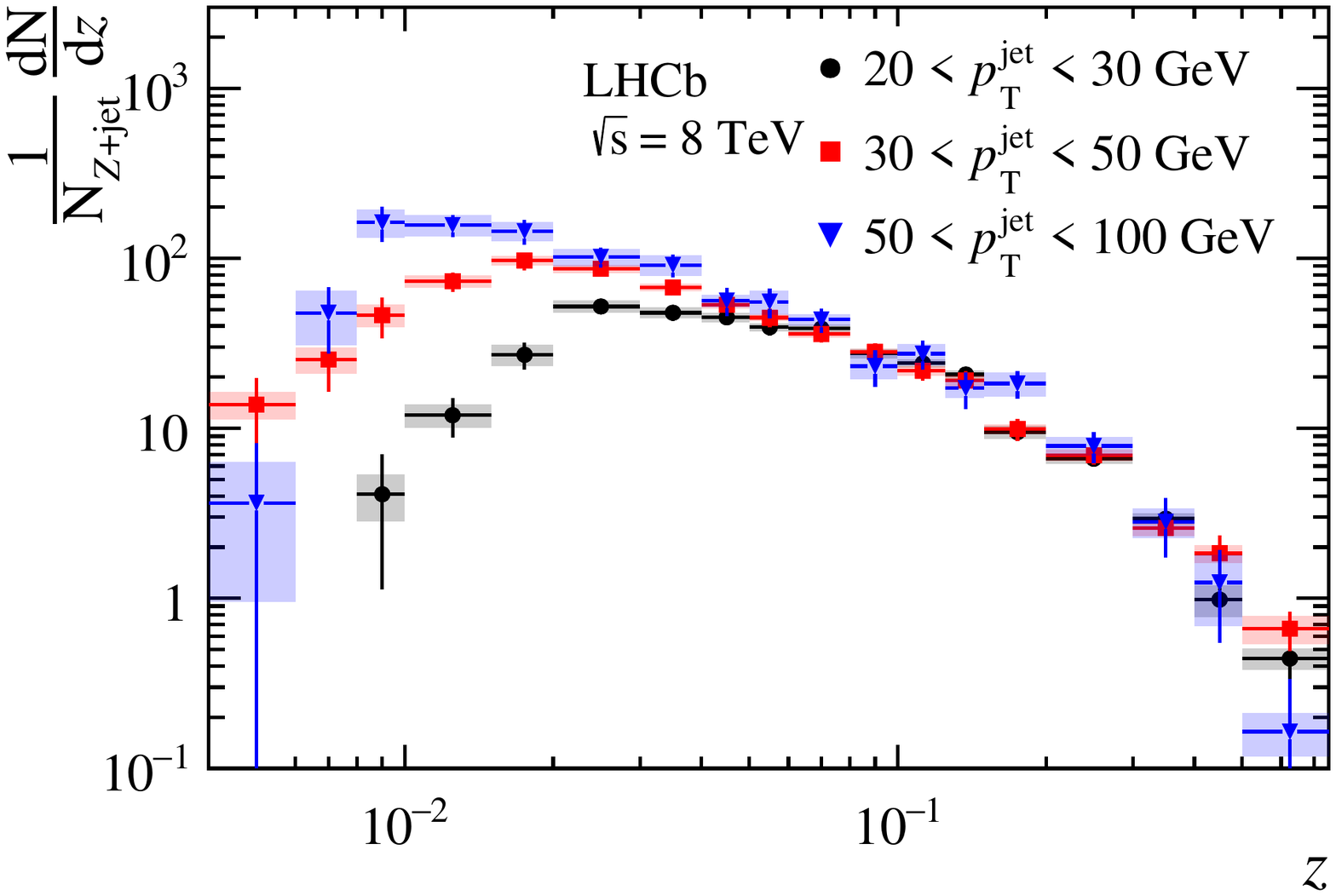}
    \caption{Distributions of the longitudinal momentum fraction of the hadron with respect to the jet in three bins of jet \pt. The bars (boxes) show the statistical (systematic) uncertainties.}
    \label{fig:zdistributions}
\end{figure}

Figure~\ref{fig:zdistributions} shows the distributions of $z$ in three jet \pt bins. These illustrate that the longitudinal momentum fraction is approximately constant as a function of jet \pt at high $z$. At low $z$ the fragmentation functions differ, which is a kinematic effect due to the requirement that the track momentum be greater than $4\gev$; therefore, higher \pt jets can probe smaller $z$. This also reflects that the charged hadron multiplicity increases with jet \pt. Comparing these forward measurements to inclusive jet measurements at central rapidity from \atlas~\cite{Aad:2011sc} indicates that the fragmentation functions are not as steeply falling at high $z$~\cite{supplemental}. This may reflect differences between light-quark and gluon fragmentation.

\begin{figure}[tb]
    \centering
    \includegraphics[width=0.6\linewidth]{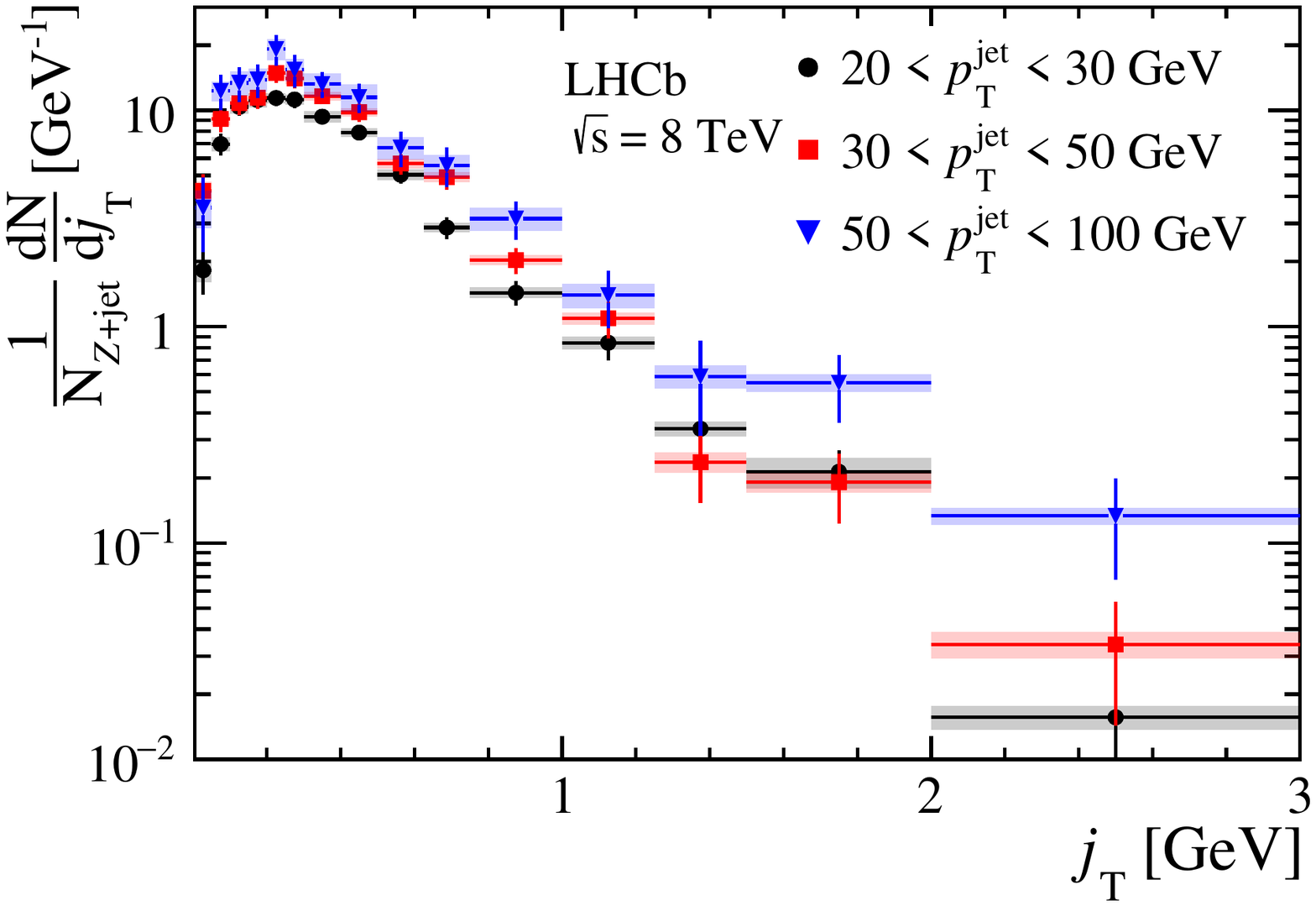}
    \caption{Distributions of  the transverse momentum of charged hadrons with respect to the jet axis in three bins of jet \pt. The bars (boxes) show the statistical (systematic) uncertainties.}
    \label{fig:jtdistributions}
\end{figure}

\begin{figure}[tb]
    \centering
    \includegraphics[width=0.6\linewidth]{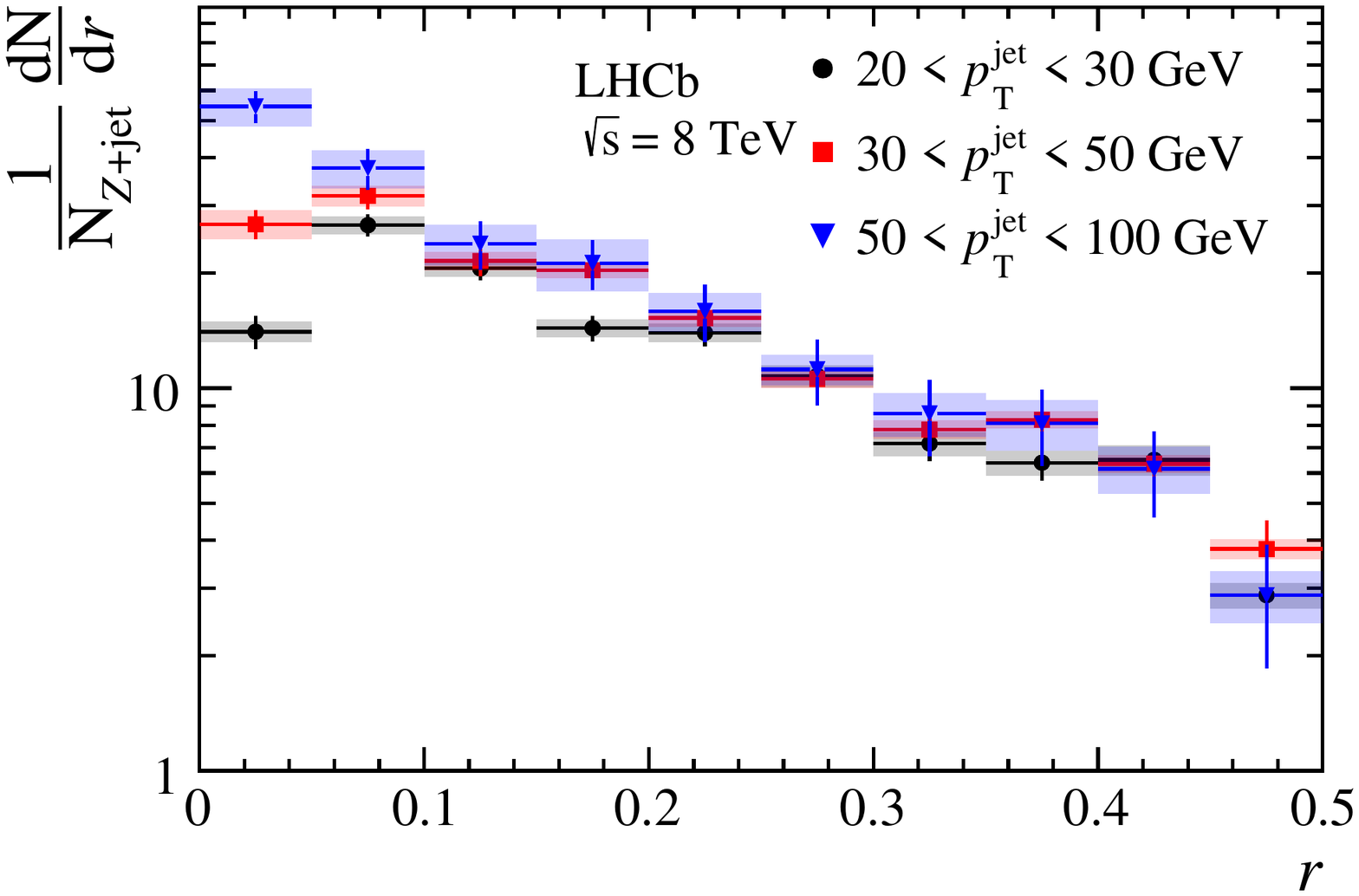}
    \caption{Radial profile distributions of hadrons with respect to the jet axis in three bins of jet \pt. The bars (boxes) show the statistical (systematic) uncertainties.}
      \label{fig:rdistributions}
\end{figure}

Figures~\ref{fig:jtdistributions} and~\ref{fig:rdistributions} show the \jt and $r$ distributions of charged hadrons within jets. The \jt profiles show a rounded peak at small \jt which transitions to a perturbative tail at larger \jt as also seen in Ref.~\cite{Acharya:2018edi}. This is indicative of an observable that can be treated in the so-called transverse-momentum-dependent framework~\cite{Collins:1981uk,Collins:1981uw,Collins:1984kg,Collins:2011zzd}, where sensitivity to both a large and small transverse momentum scale is necessary. The radial profiles show that the number of charged hadrons at small $r$ is highly dependent on jet \pt; however, the values are relatively constant as a function of jet \pt at nearly all other values of $r$. Interestingly, the \jt fragmentation distributions are similar to the central pseudorapidity inclusive jet results; however, these measurements are more collimated in $r$ than the inclusive jet measurements~\cite{supplemental}. This behavior in $r$ is correlated to the flatter fragmentation in $z$ and may be a reflection of the different pseudorapidity region or differences in light-quark and gluon fragmentation. We note that the comparisons to the measurements by \atlas should be qualitative in nature, rather than quantitative, due to the slightly different kinematic criteria placed on each of the measurements. The distributions in \jt and $r$ offer the opportunity to study the interplay between perturbative parton shower and nonperturbative hadronization dynamics. For example, the steeply falling tail of the \jt distributions results from a combination of perturbative radiation and nonperturbative hadronization processes.

%
%

%
%

\begin{figure}[tb]
    \centering
    \includegraphics[width=0.6\linewidth]{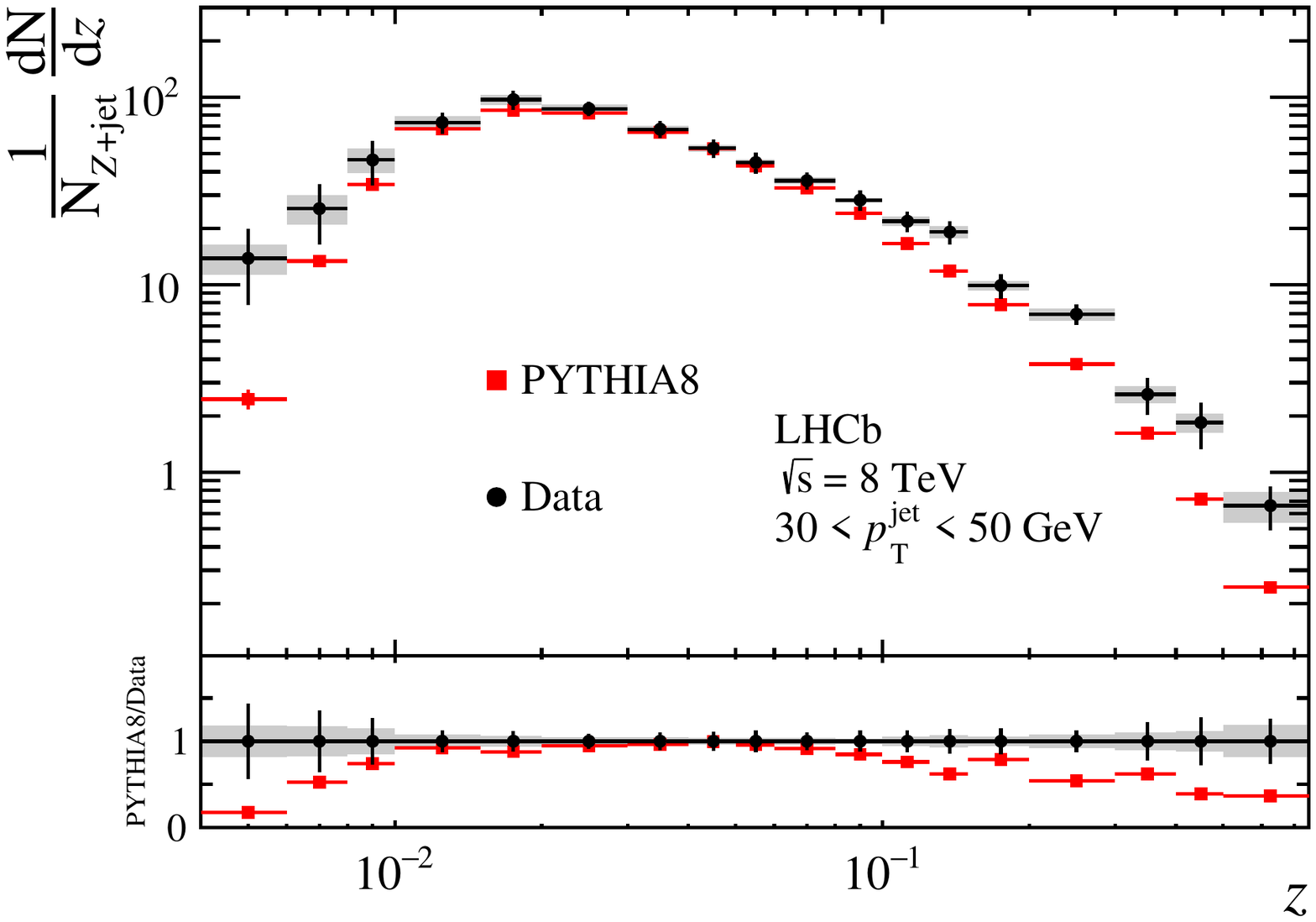}
    \caption{The $z$ distribution for jets with $30<\pt<50\gev$ compared to \pythia8 simulation predictions. The bars (boxes) show the statistical (systematic) uncertainties.}
    \label{fig:theorycomparison}
\end{figure}

The fragmentation functions are compared to predictions from \pythia8 \zjet events, where the details of the \pythia8 configuration can be found in Ref.~\cite{supplemental}. These comparisons are made since the specific LHCb tune contains realistic experimental conditions~\cite{LHCb-PROC-2010-056} and also shows that the unfolding procedure is not simply correcting the measured distributions to the predictions from \pythia8. An example of the comparison as a function of $z$ is shown in Fig.~\ref{fig:theorycomparison}; all of the comparisons described in this text can be found in Ref.~\cite{supplemental}. In general, \pythia8 underestimates the number of charged hadrons at high  $z$; \pythia8 also underestimates the number of charged hadrons at small $r$. Comparisons of the data to predictions from \pythia8 as a function of \jt show a consistent shape, but in general \pythia underestimates the number of charged hadrons in each bin by approximately 20\%. The integral of the ratio of the \pythia8 predictions to the data is always less than unity, which is a reflection of the underestimation of the mean charged hadron multiplicity in \pythia8.

In summary, the production of charged hadrons in jets recoiling against a \Z boson is measured in $\sqs=8$ \tev $pp$ collisions by the \lhcb experiment. The jets are measured in the fiducial region of $20<\pt<100\gev$ and $2.5<\eta<4$, while the hadrons are required to have $\pt>0.25\gev$, $p>4\gev$, and to be located within the jet cone of distance parameter $R=0.5$. The longitudinal momentum fraction, momentum transverse to the jet axis, and radial profile of the charged hadrons are measured with respect to the jet axis. These results provide insight into hadronization mechanisms as they probe a new kinematic regime. They additionally probe a high fraction of light-quark jets versus gluon jets when compared to midrapidity inclusive jet measurements in the same jet \pt range. The results are compared to predictions from the \pythia8 event generator with a specific LHCb configuration, and show that the simulation underestimates the number of high momentum hadrons. Additionally, comparisons with inclusive midrapidity gluon-dominated jet measurements indicate that light quark-dominated jets recoiling against a \Z boson at forward rapidity are more collimated in both $z$ and $r$~\cite{supplemental}. This work lays the foundation for a broader hadronization research program at \lhcb, utilizing the excellent tracking, particle identification, and heavy-flavor jet tagging capabilities already demonstrated by the \lhcb detector~\cite{LHCb-DP-2014-002,LHCb-PAPER-2015-016}.

\section*{Acknowledgements}

\noindent We express our gratitude to our colleagues in the CERN
accelerator departments for the excellent performance of the LHC. We
thank the technical and administrative staff at the LHCb
institutes.
We acknowledge support from CERN and from the national agencies:
CAPES, CNPq, FAPERJ and FINEP (Brazil); 
MOST and NSFC (China); 
CNRS/IN2P3 (France); 
BMBF, DFG and MPG (Germany); 
INFN (Italy); 
NWO (Netherlands); 
MNiSW and NCN (Poland); 
MEN/IFA (Romania); 
MSHE (Russia); 
MinECo (Spain); 
SNSF and SER (Switzerland); 
NASU (Ukraine); 
STFC (United Kingdom); 
NSF (USA).
We acknowledge the computing resources that are provided by CERN, IN2P3
(France), KIT and DESY (Germany), INFN (Italy), SURF (Netherlands),
PIC (Spain), GridPP (United Kingdom), RRCKI and Yandex
LLC (Russia), CSCS (Switzerland), IFIN-HH (Romania), CBPF (Brazil),
PL-GRID (Poland) and OSC (USA).
We are indebted to the communities behind the multiple open-source
software packages on which we depend.
Individual groups or members have received support from
AvH Foundation (Germany);
EPLANET, Marie Sk\l{}odowska-Curie Actions and ERC (European Union);
ANR, Labex P2IO and OCEVU, and R\'{e}gion Auvergne-Rh\^{o}ne-Alpes (France);
Key Research Program of Frontier Sciences of CAS, CAS PIFI, and the Thousand Talents Program (China);
RFBR, RSF and Yandex LLC (Russia);
GVA, XuntaGal and GENCAT (Spain);
the Royal Society
and the Leverhulme Trust (United Kingdom);
Laboratory Directed Research and Development program of LANL (USA).

\addcontentsline{toc}{section}{References}
\bibliographystyle{LHCb}
\bibliography{main,standard,LHCb-PAPER,LHCb-CONF,LHCb-DP,LHCb-TDR}

\ifx\mcitethebibliography\mciteundefinedmacro
\PackageError{LHCb.bst}{mciteplus.sty has not been loaded}
{This bibstyle requires the use of the mciteplus package.}\fi
\providecommand{\href}[2]{#2}
\begin{mcitethebibliography}{10}
\mciteSetBstSublistMode{n}
\mciteSetBstMaxWidthForm{subitem}{\alph{mcitesubitemcount})}
\mciteSetBstSublistLabelBeginEnd{\mcitemaxwidthsubitemform\space}
{\relax}{\relax}

\bibitem{Collins:1981uk}
J.~C. Collins and D.~E. Soper,
  \ifthenelse{\boolean{articletitles}}{\emph{{Back-to-back jets in QCD}},
  }{}\href{https://doi.org/10.1016/0550-3213(81)90339-4}{Nucl.\ Phys.\
  \textbf{B193} (1981) 381}, Erratum
  \href{https://doi.org/10.1016/0550-3213(81)90339-4}{ibid.\   \textbf{B213}
  (1983) 545}\relax
\mciteBstWouldAddEndPuncttrue
\mciteSetBstMidEndSepPunct{\mcitedefaultmidpunct}
{\mcitedefaultendpunct}{\mcitedefaultseppunct}\relax
\EndOfBibitem
\bibitem{Collins:1981uw}
J.~C. Collins and D.~E. Soper,
  \ifthenelse{\boolean{articletitles}}{\emph{{Parton distribution and decay
  functions}}, }{}\href{https://doi.org/10.1016/0550-3213(82)90021-9}{Nucl.\
  Phys.\  \textbf{B194} (1982) 445}\relax
\mciteBstWouldAddEndPuncttrue
\mciteSetBstMidEndSepPunct{\mcitedefaultmidpunct}
{\mcitedefaultendpunct}{\mcitedefaultseppunct}\relax
\EndOfBibitem
\bibitem{Collins:2011zzd}
J.~Collins, \ifthenelse{\boolean{articletitles}}{\emph{{Foundations of
  perturbative QCD}}, }{}Camb.\ Monogr.\ Part.\ Phys.\ Nucl.\ Phys.\ Cosmol.\
  \textbf{32} (2011) 1\relax
\mciteBstWouldAddEndPuncttrue
\mciteSetBstMidEndSepPunct{\mcitedefaultmidpunct}
{\mcitedefaultendpunct}{\mcitedefaultseppunct}\relax
\EndOfBibitem
\bibitem{Metz:2016swz}
A.~Metz and A.~Vossen, \ifthenelse{\boolean{articletitles}}{\emph{{Parton
  fragmentation functions}},
  }{}\href{https://doi.org/10.1016/j.ppnp.2016.08.003}{Prog.\ Part.\ Nucl.\
  Phys.\  \textbf{91} (2016) 136},
  \href{http://arxiv.org/abs/1607.02521}{{\normalfont\ttfamily
  arXiv:1607.02521}}\relax
\mciteBstWouldAddEndPuncttrue
\mciteSetBstMidEndSepPunct{\mcitedefaultmidpunct}
{\mcitedefaultendpunct}{\mcitedefaultseppunct}\relax
\EndOfBibitem
\bibitem{Buskulic:1994ft}
ALEPH collaboration, D.~Buskulic {\em et~al.},
  \ifthenelse{\boolean{articletitles}}{\emph{{Inclusive $\pi^\pm$, $K^\pm$ and
  ($p$,$\bar{p}$) differential cross-sections at the Z resonance}},
  }{}\href{https://doi.org/10.1007/BF01556360}{Z.\ Phys.\  \textbf{C66} (1995)
  355}\relax
\mciteBstWouldAddEndPuncttrue
\mciteSetBstMidEndSepPunct{\mcitedefaultmidpunct}
{\mcitedefaultendpunct}{\mcitedefaultseppunct}\relax
\EndOfBibitem
\bibitem{Akers:1994ez}
OPAL collaboration, R.~Akers {\em et~al.},
  \ifthenelse{\boolean{articletitles}}{\emph{{Measurement of the production
  rates of charged hadrons in $e^+e^-$ annihilation at the $Z^0$}},
  }{}\href{https://doi.org/10.1007/BF01411010}{Z.\ Phys.\  \textbf{C63} (1994)
  181}\relax
\mciteBstWouldAddEndPuncttrue
\mciteSetBstMidEndSepPunct{\mcitedefaultmidpunct}
{\mcitedefaultendpunct}{\mcitedefaultseppunct}\relax
\EndOfBibitem
\bibitem{OPAL:1995ab}
OPAL collaboration, R.~Akers {\em et~al.},
  \ifthenelse{\boolean{articletitles}}{\emph{{A model independent measurement
  of quark and gluon jet properties and differences}},
  }{}\href{https://doi.org/10.1007/BF01566667}{Z.\ Phys.\  \textbf{C68} (1995)
  179}\relax
\mciteBstWouldAddEndPuncttrue
\mciteSetBstMidEndSepPunct{\mcitedefaultmidpunct}
{\mcitedefaultendpunct}{\mcitedefaultseppunct}\relax
\EndOfBibitem
\bibitem{Abreu:1998vq}
DELPHI collaboration, P.~Abreu {\em et~al.},
  \ifthenelse{\boolean{articletitles}}{\emph{{$\pi^\pm$, $K^\pm$, $p$ and
  $\bar{p}$ production in $Z^0\rightarrow q\bar{q}$, $Z^0\rightarrow b\bar{b}$,
  $Z^0\rightarrow u\bar{u}, d\bar{d}, s\bar{s}$}},
  }{}\href{https://doi.org/10.1007/s100529800989}{Eur.\ Phys.\ J.\  \textbf{C5}
  (1998) 585}\relax
\mciteBstWouldAddEndPuncttrue
\mciteSetBstMidEndSepPunct{\mcitedefaultmidpunct}
{\mcitedefaultendpunct}{\mcitedefaultseppunct}\relax
\EndOfBibitem
\bibitem{Abe:1998zs}
SLD collaboration, K.~Abe {\em et~al.},
  \ifthenelse{\boolean{articletitles}}{\emph{{Production of $\pi^+$, $K^+$,
  $K^0$, $K^{*0}$, $\phi$, $p$ and $\Lambdares^0$ in hadronic $Z^0$ decays}},
  }{}\href{https://doi.org/10.1103/PhysRevD.59.052001}{Phys.\ Rev.\
  \textbf{D59} (1999) 052001},
  \href{http://arxiv.org/abs/hep-ex/9805029}{{\normalfont\ttfamily
  arXiv:hep-ex/9805029}}\relax
\mciteBstWouldAddEndPuncttrue
\mciteSetBstMidEndSepPunct{\mcitedefaultmidpunct}
{\mcitedefaultendpunct}{\mcitedefaultseppunct}\relax
\EndOfBibitem
\bibitem{Abreu:1999af}
DELPHI collaboration, P.~Abreu {\em et~al.},
  \ifthenelse{\boolean{articletitles}}{\emph{{Measurement of the gluon
  fragmentation function and a comparison of the scaling violation in gluon and
  quark jets}}, }{}\href{https://doi.org/10.1007/s100520050719}{Eur.\ Phys.\
  J.\  \textbf{C13} (2000) 573}\relax
\mciteBstWouldAddEndPuncttrue
\mciteSetBstMidEndSepPunct{\mcitedefaultmidpunct}
{\mcitedefaultendpunct}{\mcitedefaultseppunct}\relax
\EndOfBibitem
\bibitem{Lees:2013rqd}
BaBar collaboration, J.~P. Lees {\em et~al.},
  \ifthenelse{\boolean{articletitles}}{\emph{{Production of charged pions,
  kaons, and protons in $e^+e^-$ annihilations into hadrons at $\sqrt{s}$=10.54
  \gev}}, }{}\href{https://doi.org/10.1103/PhysRevD.88.032011}{Phys.\ Rev.\
  \textbf{D88} (2013) 032011},
  \href{http://arxiv.org/abs/1306.2895}{{\normalfont\ttfamily
  arXiv:1306.2895}}\relax
\mciteBstWouldAddEndPuncttrue
\mciteSetBstMidEndSepPunct{\mcitedefaultmidpunct}
{\mcitedefaultendpunct}{\mcitedefaultseppunct}\relax
\EndOfBibitem
\bibitem{Leitgab:2013qh}
Belle collaboration, M.~Leitgab {\em et~al.},
  \ifthenelse{\boolean{articletitles}}{\emph{{Precision measurement of charged
  pion and kaon differential cross sections in $e^+e^-$ annihilation at
  $\sqrt{s}$=10.52 \gev}},
  }{}\href{https://doi.org/10.1103/PhysRevLett.111.062002}{Phys.\ Rev.\ Lett.\
  \textbf{111} (2013) 062002},
  \href{http://arxiv.org/abs/1301.6183}{{\normalfont\ttfamily
  arXiv:1301.6183}}\relax
\mciteBstWouldAddEndPuncttrue
\mciteSetBstMidEndSepPunct{\mcitedefaultmidpunct}
{\mcitedefaultendpunct}{\mcitedefaultseppunct}\relax
\EndOfBibitem
\bibitem{Seidl:2019jei}
Belle collaboration, R.~Seidl {\em et~al.},
  \ifthenelse{\boolean{articletitles}}{\emph{{Transverse momentum dependent
  production cross sections of charged pions, kaons and protons produced in
  inclusive $e^+e^-$ annihilation at $\sqrt{s}=$ 10.58 \gev}},
  }{}\href{http://arxiv.org/abs/1902.01552}{{\normalfont\ttfamily
  arXiv:1902.01552}}\relax
\mciteBstWouldAddEndPuncttrue
\mciteSetBstMidEndSepPunct{\mcitedefaultmidpunct}
{\mcitedefaultendpunct}{\mcitedefaultseppunct}\relax
\EndOfBibitem
\bibitem{Airapetian:2012ki}
HERMES collaboration, A.~Airapetian {\em et~al.},
  \ifthenelse{\boolean{articletitles}}{\emph{{Multiplicities of charged pions
  and kaons from semi-inclusive deep-inelastic scattering by the proton and the
  deuteron}}, }{}\href{https://doi.org/10.1103/PhysRevD.87.074029}{Phys.\ Rev.\
   \textbf{D87} (2013) 074029},
  \href{http://arxiv.org/abs/1212.5407}{{\normalfont\ttfamily
  arXiv:1212.5407}}\relax
\mciteBstWouldAddEndPuncttrue
\mciteSetBstMidEndSepPunct{\mcitedefaultmidpunct}
{\mcitedefaultendpunct}{\mcitedefaultseppunct}\relax
\EndOfBibitem
\bibitem{Aghasyan:2017ctw}
COMPASS collaboration, M.~Aghasyan {\em et~al.},
  \ifthenelse{\boolean{articletitles}}{\emph{{Transverse-momentum-dependent
  multiplicities of charged hadrons in muon-deuteron deep inelastic
  scattering}}, }{}\href{https://doi.org/10.1103/PhysRevD.97.032006}{Phys.\
  Rev.\  \textbf{D97} (2018) 032006},
  \href{http://arxiv.org/abs/1709.07374}{{\normalfont\ttfamily
  arXiv:1709.07374}}\relax
\mciteBstWouldAddEndPuncttrue
\mciteSetBstMidEndSepPunct{\mcitedefaultmidpunct}
{\mcitedefaultendpunct}{\mcitedefaultseppunct}\relax
\EndOfBibitem
\bibitem{Adams:2006nd}
STAR collaboration, J.~Adams {\em et~al.},
  \ifthenelse{\boolean{articletitles}}{\emph{{Identified hadron spectra at
  large transverse momentum in p+p and d+Au collisions at \sqsnn = 200 \gev}},
  }{}\href{https://doi.org/10.1016/j.physletb.2006.04.032}{Phys.\ Lett.\
  \textbf{B637} (2006) 161},
  \href{http://arxiv.org/abs/nucl-ex/0601033}{{\normalfont\ttfamily
  arXiv:nucl-ex/0601033}}\relax
\mciteBstWouldAddEndPuncttrue
\mciteSetBstMidEndSepPunct{\mcitedefaultmidpunct}
{\mcitedefaultendpunct}{\mcitedefaultseppunct}\relax
\EndOfBibitem
\bibitem{Arsene:2007jd}
BRAHMS collaboration, I.~Arsene {\em et~al.},
  \ifthenelse{\boolean{articletitles}}{\emph{{Production of mesons and baryons
  at high rapidity and high $p_T$ in proton-proton collisions at $\sqrt{s}$ =
  200 \gev}}, }{}\href{https://doi.org/10.1103/PhysRevLett.98.252001}{Phys.\
  Rev.\ Lett.\  \textbf{98} (2007) 252001},
  \href{http://arxiv.org/abs/hep-ex/0701041}{{\normalfont\ttfamily
  arXiv:hep-ex/0701041}}\relax
\mciteBstWouldAddEndPuncttrue
\mciteSetBstMidEndSepPunct{\mcitedefaultmidpunct}
{\mcitedefaultendpunct}{\mcitedefaultseppunct}\relax
\EndOfBibitem
\bibitem{Adare:2014wht}
PHENIX collaboration, A.~Adare {\em et~al.},
  \ifthenelse{\boolean{articletitles}}{\emph{{Charged-pion cross sections and
  double-helicity asymmetries in polarized p+p collisions at $\sqrt{s}$=200
  \gev}}, }{}\href{https://doi.org/10.1103/PhysRevD.91.032001}{Phys.\ Rev.\
  \textbf{D91} (2015) 032001},
  \href{http://arxiv.org/abs/1409.1907}{{\normalfont\ttfamily
  arXiv:1409.1907}}\relax
\mciteBstWouldAddEndPuncttrue
\mciteSetBstMidEndSepPunct{\mcitedefaultmidpunct}
{\mcitedefaultendpunct}{\mcitedefaultseppunct}\relax
\EndOfBibitem
\bibitem{Collins:2002kn}
J.~C. Collins, \ifthenelse{\boolean{articletitles}}{\emph{{Leading-twist
  single-transverse-spin asymmetries: Drell-Yan and deep-inelastic
  scattering}}, }{}\href{https://doi.org/10.1016/S0370-2693(02)01819-1}{Phys.\
  Lett.\  \textbf{B536} (2002) 43},
  \href{http://arxiv.org/abs/hep-ph/0204004}{{\normalfont\ttfamily
  arXiv:hep-ph/0204004}}\relax
\mciteBstWouldAddEndPuncttrue
\mciteSetBstMidEndSepPunct{\mcitedefaultmidpunct}
{\mcitedefaultendpunct}{\mcitedefaultseppunct}\relax
\EndOfBibitem
\bibitem{Rogers:2010dm}
T.~C. Rogers and P.~J. Mulders, \ifthenelse{\boolean{articletitles}}{\emph{{No
  generalized transverse momentum dependent factorization in the
  hadroproduction of high transverse momentum hadrons}},
  }{}\href{https://doi.org/10.1103/PhysRevD.81.094006}{Phys.\ Rev.\
  \textbf{D81} (2010) 094006},
  \href{http://arxiv.org/abs/1001.2977}{{\normalfont\ttfamily
  arXiv:1001.2977}}\relax
\mciteBstWouldAddEndPuncttrue
\mciteSetBstMidEndSepPunct{\mcitedefaultmidpunct}
{\mcitedefaultendpunct}{\mcitedefaultseppunct}\relax
\EndOfBibitem
\bibitem{Cacciari:2008gp}
M.~Cacciari, G.~P. Salam, and G.~Soyez,
  \ifthenelse{\boolean{articletitles}}{\emph{{The anti-$k_t$ jet clustering
  algorithm}}, }{}\href{https://doi.org/10.1088/1126-6708/2008/04/063}{JHEP
  \textbf{04} (2008) 063},
  \href{http://arxiv.org/abs/0802.1189}{{\normalfont\ttfamily
  arXiv:0802.1189}}\relax
\mciteBstWouldAddEndPuncttrue
\mciteSetBstMidEndSepPunct{\mcitedefaultmidpunct}
{\mcitedefaultendpunct}{\mcitedefaultseppunct}\relax
\EndOfBibitem
\bibitem{Sjostrand:2006za}
T.~Sj\"{o}strand, S.~Mrenna, and P.~Skands,
  \ifthenelse{\boolean{articletitles}}{\emph{{PYTHIA 6.4 physics and manual}},
  }{}\href{https://doi.org/10.1088/1126-6708/2006/05/026}{JHEP \textbf{05}
  (2006) 026}, \href{http://arxiv.org/abs/hep-ph/0603175}{{\normalfont\ttfamily
  arXiv:hep-ph/0603175}}\relax
\mciteBstWouldAddEndPuncttrue
\mciteSetBstMidEndSepPunct{\mcitedefaultmidpunct}
{\mcitedefaultendpunct}{\mcitedefaultseppunct}\relax
\EndOfBibitem
\bibitem{Sjostrand:2007gs}
T.~Sj\"{o}strand, S.~Mrenna, and P.~Skands,
  \ifthenelse{\boolean{articletitles}}{\emph{{A brief introduction to PYTHIA
  8.1}}, }{}\href{https://doi.org/10.1016/j.cpc.2008.01.036}{Comput.\ Phys.\
  Commun.\  \textbf{178} (2008) 852},
  \href{http://arxiv.org/abs/0710.3820}{{\normalfont\ttfamily
  arXiv:0710.3820}}\relax
\mciteBstWouldAddEndPuncttrue
\mciteSetBstMidEndSepPunct{\mcitedefaultmidpunct}
{\mcitedefaultendpunct}{\mcitedefaultseppunct}\relax
\EndOfBibitem
\bibitem{Aad:2011sc}
ATLAS collaboration, G.~Aad {\em et~al.},
  \ifthenelse{\boolean{articletitles}}{\emph{{Measurement of the jet
  fragmentation function and transverse profile in proton-proton collisions at
  a center-of-mass energy of 7 \tev with the ATLAS detector}},
  }{}\href{https://doi.org/10.1140/epjc/s10052-011-1795-y}{Eur.\ Phys.\ J.\
  \textbf{C71} (2011) 1795},
  \href{http://arxiv.org/abs/1109.5816}{{\normalfont\ttfamily
  arXiv:1109.5816}}\relax
\mciteBstWouldAddEndPuncttrue
\mciteSetBstMidEndSepPunct{\mcitedefaultmidpunct}
{\mcitedefaultendpunct}{\mcitedefaultseppunct}\relax
\EndOfBibitem
\bibitem{Chatrchyan:2014ava}
CMS collaboration, S.~Chatrchyan {\em et~al.},
  \ifthenelse{\boolean{articletitles}}{\emph{{Measurement of jet fragmentation
  in PbPb and pp collisions at $\sqsnn=2.76$ \tev}},
  }{}\href{https://doi.org/10.1103/PhysRevC.90.024908}{Phys.\ Rev.\
  \textbf{C90} (2014) 024908},
  \href{http://arxiv.org/abs/1406.0932}{{\normalfont\ttfamily
  arXiv:1406.0932}}\relax
\mciteBstWouldAddEndPuncttrue
\mciteSetBstMidEndSepPunct{\mcitedefaultmidpunct}
{\mcitedefaultendpunct}{\mcitedefaultseppunct}\relax
\EndOfBibitem
\bibitem{ALICE:2014dla}
ALICE collaboration, B.~Abelev {\em et~al.},
  \ifthenelse{\boolean{articletitles}}{\emph{{Charged jet cross sections and
  properties in proton-proton collisions at $\sqrt{s}=7$ \tev}},
  }{}\href{https://doi.org/10.1103/PhysRevD.91.112012}{Phys.\ Rev.\
  \textbf{D91} (2015) 112012},
  \href{http://arxiv.org/abs/1411.4969}{{\normalfont\ttfamily
  arXiv:1411.4969}}\relax
\mciteBstWouldAddEndPuncttrue
\mciteSetBstMidEndSepPunct{\mcitedefaultmidpunct}
{\mcitedefaultendpunct}{\mcitedefaultseppunct}\relax
\EndOfBibitem
\bibitem{Aaboud:2017tke}
ATLAS collaboration, M.~Aaboud {\em et~al.},
  \ifthenelse{\boolean{articletitles}}{\emph{{Measurement of jet fragmentation
  in 5.02 \tev proton-lead and proton-proton collisions with the ATLAS
  detector}}, }{}\href{https://doi.org/10.1016/j.nuclphysa.2018.07.006}{Nucl.\
  Phys.\  \textbf{A978} (2018) 65},
  \href{http://arxiv.org/abs/1706.02859}{{\normalfont\ttfamily
  arXiv:1706.02859}}\relax
\mciteBstWouldAddEndPuncttrue
\mciteSetBstMidEndSepPunct{\mcitedefaultmidpunct}
{\mcitedefaultendpunct}{\mcitedefaultseppunct}\relax
\EndOfBibitem
\bibitem{Aaboud:2017qwh}
ATLAS collaboration, M.~Aaboud {\em et~al.},
  \ifthenelse{\boolean{articletitles}}{\emph{{Measurement of the soft-drop jet
  mass in pp collisions at $\sqrt{s} = 13$ \tev with the ATLAS detector}},
  }{}\href{https://doi.org/10.1103/PhysRevLett.121.092001}{Phys.\ Rev.\ Lett.\
  \textbf{121} (2018) 092001},
  \href{http://arxiv.org/abs/1711.08341}{{\normalfont\ttfamily
  arXiv:1711.08341}}\relax
\mciteBstWouldAddEndPuncttrue
\mciteSetBstMidEndSepPunct{\mcitedefaultmidpunct}
{\mcitedefaultendpunct}{\mcitedefaultseppunct}\relax
\EndOfBibitem
\bibitem{Acharya:2018eat}
ALICE collaboration, S.~Acharya {\em et~al.},
  \ifthenelse{\boolean{articletitles}}{\emph{{Charged jet cross section and
  fragmentation in proton-proton collisions at $\sqrt{s}$ = 7 \tev}},
  }{}\href{https://doi.org/10.1103/PhysRevD.99.012016}{Phys.\ Rev.\
  \textbf{D99} (2019) 012016},
  \href{http://arxiv.org/abs/1809.03232}{{\normalfont\ttfamily
  arXiv:1809.03232}}\relax
\mciteBstWouldAddEndPuncttrue
\mciteSetBstMidEndSepPunct{\mcitedefaultmidpunct}
{\mcitedefaultendpunct}{\mcitedefaultseppunct}\relax
\EndOfBibitem
\bibitem{Aaboud:2019oac}
ATLAS collaboration, M.~Aaboud {\em et~al.},
  \ifthenelse{\boolean{articletitles}}{\emph{{Comparison of Fragmentation
  Functions for Jets Dominated by Light Quarks and Gluons from $pp$ and Pb+Pb
  Collisions in ATLAS}},
  }{}\href{https://doi.org/10.1103/PhysRevLett.123.042001}{Phys.\ Rev.\ Lett.\
  \textbf{123} (2019) 042001},
  \href{http://arxiv.org/abs/1902.10007}{{\normalfont\ttfamily
  arXiv:1902.10007}}\relax
\mciteBstWouldAddEndPuncttrue
\mciteSetBstMidEndSepPunct{\mcitedefaultmidpunct}
{\mcitedefaultendpunct}{\mcitedefaultseppunct}\relax
\EndOfBibitem
\bibitem{Alves:2008zz}
LHCb collaboration, A.~A. Alves~Jr.\ {\em et~al.},
  \ifthenelse{\boolean{articletitles}}{\emph{{The \lhcb detector at the LHC}},
  }{}\href{https://doi.org/10.1088/1748-0221/3/08/S08005}{JINST \textbf{3}
  (2008) S08005}\relax
\mciteBstWouldAddEndPuncttrue
\mciteSetBstMidEndSepPunct{\mcitedefaultmidpunct}
{\mcitedefaultendpunct}{\mcitedefaultseppunct}\relax
\EndOfBibitem
\bibitem{LHCb-DP-2014-002}
LHCb collaboration, R.~Aaij {\em et~al.},
  \ifthenelse{\boolean{articletitles}}{\emph{{LHCb detector performance}},
  }{}\href{https://doi.org/10.1142/S0217751X15300227}{Int.\ J.\ Mod.\ Phys.\
  \textbf{A30} (2015) 1530022},
  \href{http://arxiv.org/abs/1412.6352}{{\normalfont\ttfamily
  arXiv:1412.6352}}\relax
\mciteBstWouldAddEndPuncttrue
\mciteSetBstMidEndSepPunct{\mcitedefaultmidpunct}
{\mcitedefaultendpunct}{\mcitedefaultseppunct}\relax
\EndOfBibitem
\bibitem{LHCb-PROC-2010-056}
I.~Belyaev {\em et~al.}, \ifthenelse{\boolean{articletitles}}{\emph{{Handling
  of the generation of primary events in Gauss, the LHCb simulation
  framework}}, }{}\href{https://doi.org/10.1088/1742-6596/331/3/032047}{J.\
  Phys.\ Conf.\ Ser.\  \textbf{331} (2011) 032047}\relax
\mciteBstWouldAddEndPuncttrue
\mciteSetBstMidEndSepPunct{\mcitedefaultmidpunct}
{\mcitedefaultendpunct}{\mcitedefaultseppunct}\relax
\EndOfBibitem
\bibitem{Lange:2001uf}
D.~J. Lange, \ifthenelse{\boolean{articletitles}}{\emph{{The EvtGen particle
  decay simulation package}},
  }{}\href{https://doi.org/10.1016/S0168-9002(01)00089-4}{Nucl.\ Instrum.\
  Meth.\  \textbf{A462} (2001) 152}\relax
\mciteBstWouldAddEndPuncttrue
\mciteSetBstMidEndSepPunct{\mcitedefaultmidpunct}
{\mcitedefaultendpunct}{\mcitedefaultseppunct}\relax
\EndOfBibitem
\bibitem{Golonka:2005pn}
P.~Golonka and Z.~Was, \ifthenelse{\boolean{articletitles}}{\emph{{PHOTOS Monte
  Carlo: A precision tool for QED corrections in $Z$ and $W$ decays}},
  }{}\href{https://doi.org/10.1140/epjc/s2005-02396-4}{Eur.\ Phys.\ J.\
  \textbf{C45} (2006) 97},
  \href{http://arxiv.org/abs/hep-ph/0506026}{{\normalfont\ttfamily
  arXiv:hep-ph/0506026}}\relax
\mciteBstWouldAddEndPuncttrue
\mciteSetBstMidEndSepPunct{\mcitedefaultmidpunct}
{\mcitedefaultendpunct}{\mcitedefaultseppunct}\relax
\EndOfBibitem
\bibitem{Agostinelli:2002hh}
Geant4 collaboration, S.~Agostinelli {\em et~al.},
  \ifthenelse{\boolean{articletitles}}{\emph{{Geant4: A simulation toolkit}},
  }{}\href{https://doi.org/10.1016/S0168-9002(03)01368-8}{Nucl.\ Instrum.\
  Meth.\  \textbf{A506} (2003) 250}\relax
\mciteBstWouldAddEndPuncttrue
\mciteSetBstMidEndSepPunct{\mcitedefaultmidpunct}
{\mcitedefaultendpunct}{\mcitedefaultseppunct}\relax
\EndOfBibitem
\bibitem{Allison:2006ve}
Geant4 collaboration, J.~Allison {\em et~al.},
  \ifthenelse{\boolean{articletitles}}{\emph{{Geant4 developments and
  applications}}, }{}\href{https://doi.org/10.1109/TNS.2006.869826}{IEEE
  Trans.\ Nucl.\ Sci.\  \textbf{53} (2006) 270}\relax
\mciteBstWouldAddEndPuncttrue
\mciteSetBstMidEndSepPunct{\mcitedefaultmidpunct}
{\mcitedefaultendpunct}{\mcitedefaultseppunct}\relax
\EndOfBibitem
\bibitem{LHCb-PROC-2011-006}
M.~Clemencic {\em et~al.}, \ifthenelse{\boolean{articletitles}}{\emph{{The
  \lhcb simulation application, Gauss: Design, evolution and experience}},
  }{}\href{https://doi.org/10.1088/1742-6596/331/3/032023}{J.\ Phys.\ Conf.\
  Ser.\  \textbf{331} (2011) 032023}\relax
\mciteBstWouldAddEndPuncttrue
\mciteSetBstMidEndSepPunct{\mcitedefaultmidpunct}
{\mcitedefaultendpunct}{\mcitedefaultseppunct}\relax
\EndOfBibitem
\bibitem{LHCb-PAPER-2016-011}
LHCb collaboration, R.~Aaij {\em et~al.},
  \ifthenelse{\boolean{articletitles}}{\emph{{Measurement of forward $\W\!$ and
  $\Z$ boson production in association with jets in proton-proton collisions at
  $\sqrt{s}=8$\,TeV}}, }{}\href{https://doi.org/10.1007/JHEP05(2016)131}{JHEP
  \textbf{05} (2016) 131},
  \href{http://arxiv.org/abs/1605.00951}{{\normalfont\ttfamily
  arXiv:1605.00951}}\relax
\mciteBstWouldAddEndPuncttrue
\mciteSetBstMidEndSepPunct{\mcitedefaultmidpunct}
{\mcitedefaultendpunct}{\mcitedefaultseppunct}\relax
\EndOfBibitem
\bibitem{LHCb-DP-2012-004}
R.~Aaij {\em et~al.}, \ifthenelse{\boolean{articletitles}}{\emph{{The \lhcb
  trigger and its performance in 2011}},
  }{}\href{https://doi.org/10.1088/1748-0221/8/04/P04022}{JINST \textbf{8}
  (2013) P04022}, \href{http://arxiv.org/abs/1211.3055}{{\normalfont\ttfamily
  arXiv:1211.3055}}\relax
\mciteBstWouldAddEndPuncttrue
\mciteSetBstMidEndSepPunct{\mcitedefaultmidpunct}
{\mcitedefaultendpunct}{\mcitedefaultseppunct}\relax
\EndOfBibitem
\bibitem{LHCb-PAPER-2015-049}
LHCb collaboration, R.~Aaij {\em et~al.},
  \ifthenelse{\boolean{articletitles}}{\emph{{Measurement of forward $\W\!$ and
  $\Z$ boson production in $\proton\proton$ collisions at $\sqrt{s}=8$\,TeV }},
  }{}\href{https://doi.org/10.1007/JHEP01(2016)155}{JHEP \textbf{01} (2016)
  155}, \href{http://arxiv.org/abs/1511.08039}{{\normalfont\ttfamily
  arXiv:1511.08039}}\relax
\mciteBstWouldAddEndPuncttrue
\mciteSetBstMidEndSepPunct{\mcitedefaultmidpunct}
{\mcitedefaultendpunct}{\mcitedefaultseppunct}\relax
\EndOfBibitem
\bibitem{LHCb-PAPER-2013-058}
LHCb collaboration, R.~Aaij {\em et~al.},
  \ifthenelse{\boolean{articletitles}}{\emph{{Study of forward $\Z$+jet
  production in $\proton\proton$ collisions at $\sqrt{s} = 7$\tev}},
  }{}\href{https://doi.org/10.1007/JHEP01(2014)033}{JHEP \textbf{01} (2014)
  033}, \href{http://arxiv.org/abs/1310.8197}{{\normalfont\ttfamily
  arXiv:1310.8197}}\relax
\mciteBstWouldAddEndPuncttrue
\mciteSetBstMidEndSepPunct{\mcitedefaultmidpunct}
{\mcitedefaultendpunct}{\mcitedefaultseppunct}\relax
\EndOfBibitem
\bibitem{Cacciari:2011ma}
M.~Cacciari, G.~P. Salam, and G.~Soyez,
  \ifthenelse{\boolean{articletitles}}{\emph{{FastJet user manual}},
  }{}\href{https://doi.org/10.1140/epjc/s10052-012-1896-2}{Eur.\ Phys.\ J.\
  \textbf{C72} (2012) 1896},
  \href{http://arxiv.org/abs/1111.6097}{{\normalfont\ttfamily
  arXiv:1111.6097}}\relax
\mciteBstWouldAddEndPuncttrue
\mciteSetBstMidEndSepPunct{\mcitedefaultmidpunct}
{\mcitedefaultendpunct}{\mcitedefaultseppunct}\relax
\EndOfBibitem
\bibitem{LHCb-PAPER-2016-064}
LHCb collaboration, R.~Aaij {\em et~al.},
  \ifthenelse{\boolean{articletitles}}{\emph{{Study of \jpsi production in
  jets}}, }{}\href{https://doi.org/10.1103/PhysRevLett.118.192001}{Phys.\ Rev.\
  Lett.\  \textbf{118} (2017) 192001},
  \href{http://arxiv.org/abs/1701.05116}{{\normalfont\ttfamily
  arXiv:1701.05116}}\relax
\mciteBstWouldAddEndPuncttrue
\mciteSetBstMidEndSepPunct{\mcitedefaultmidpunct}
{\mcitedefaultendpunct}{\mcitedefaultseppunct}\relax
\EndOfBibitem
\bibitem{LHCb-PAPER-2015-001}
LHCb collaboration, R.~Aaij {\em et~al.},
  \ifthenelse{\boolean{articletitles}}{\emph{{Measurement of the forward $\Z$
  boson cross-section in $\proton\proton$ collisions at $\sqrt{s}=7$\tev}},
  }{}\href{https://doi.org/10.1007/JHEP08(2015)039}{JHEP \textbf{08} (2015)
  039}, \href{http://arxiv.org/abs/1505.07024}{{\normalfont\ttfamily
  arXiv:1505.07024}}\relax
\mciteBstWouldAddEndPuncttrue
\mciteSetBstMidEndSepPunct{\mcitedefaultmidpunct}
{\mcitedefaultendpunct}{\mcitedefaultseppunct}\relax
\EndOfBibitem
\bibitem{DAgostini:1994fjx}
G.~D'Agostini, \ifthenelse{\boolean{articletitles}}{\emph{{A multidimensional
  unfolding method based on Bayes' theorem}},
  }{}\href{https://doi.org/10.1016/0168-9002(95)00274-X}{Nucl.\ Instrum.\
  Meth.\  \textbf{A362} (1995) 487}\relax
\mciteBstWouldAddEndPuncttrue
\mciteSetBstMidEndSepPunct{\mcitedefaultmidpunct}
{\mcitedefaultendpunct}{\mcitedefaultseppunct}\relax
\EndOfBibitem
\bibitem{Adye:2011gm}
T.~Adye, \ifthenelse{\boolean{articletitles}}{\emph{{Unfolding algorithms and
  tests using RooUnfold}},
  }{}\href{http://arxiv.org/abs/1105.1160}{{\normalfont\ttfamily
  arXiv:1105.1160}}\relax
\mciteBstWouldAddEndPuncttrue
\mciteSetBstMidEndSepPunct{\mcitedefaultmidpunct}
{\mcitedefaultendpunct}{\mcitedefaultseppunct}\relax
\EndOfBibitem
\bibitem{supplemental}
{See Supplemental Material for further details}\relax
\mciteBstWouldAddEndPuncttrue
\mciteSetBstMidEndSepPunct{\mcitedefaultmidpunct}
{\mcitedefaultendpunct}{\mcitedefaultseppunct}\relax
\EndOfBibitem
\bibitem{Acharya:2018edi}
ALICE collaboration, S.~Acharya {\em et~al.},
  \ifthenelse{\boolean{articletitles}}{\emph{{Jet fragmentation transverse
  momentum measurements from di-hadron correlations in $\sqrt{s}$ = 7 TeV pp
  and $\sqrt{s_{\rm{NN}}}$ = 5.02 TeV p-Pb collisions}},
  }{}\href{https://doi.org/10.1007/JHEP03(2019)169}{JHEP \textbf{03} (2019)
  169}, \href{http://arxiv.org/abs/1811.09742}{{\normalfont\ttfamily
  arXiv:1811.09742}}\relax
\mciteBstWouldAddEndPuncttrue
\mciteSetBstMidEndSepPunct{\mcitedefaultmidpunct}
{\mcitedefaultendpunct}{\mcitedefaultseppunct}\relax
\EndOfBibitem
\bibitem{Collins:1984kg}
J.~C. Collins, D.~E. Soper, and G.~F. Sterman,
  \ifthenelse{\boolean{articletitles}}{\emph{{Transverse momentum distribution
  in Drell-Yan pair and W and Z boson production}},
  }{}\href{https://doi.org/10.1016/0550-3213(85)90479-1}{Nucl.\ Phys.\
  \textbf{B250} (1985) 199}\relax
\mciteBstWouldAddEndPuncttrue
\mciteSetBstMidEndSepPunct{\mcitedefaultmidpunct}
{\mcitedefaultendpunct}{\mcitedefaultseppunct}\relax
\EndOfBibitem
\bibitem{LHCb-PAPER-2015-016}
LHCb collaboration, R.~Aaij {\em et~al.},
  \ifthenelse{\boolean{articletitles}}{\emph{{Identification of beauty and
  charm quark jets at LHCb}},
  }{}\href{https://doi.org/10.1088/1748-0221/10/06/P06013}{JINST \textbf{10}
  (2015) P06013}, \href{http://arxiv.org/abs/1504.07670}{{\normalfont\ttfamily
  arXiv:1504.07670}}\relax
\mciteBstWouldAddEndPuncttrue
\mciteSetBstMidEndSepPunct{\mcitedefaultmidpunct}
{\mcitedefaultendpunct}{\mcitedefaultseppunct}\relax
\EndOfBibitem
\end{mcitethebibliography}

\newpage
\clearpage


\centerline
{\large\bf LHCb Collaboration}
\begin
{flushleft}
\small
R.~Aaij$^{29}$,
C.~Abell{\'a}n~Beteta$^{46}$,
B.~Adeva$^{43}$,
M.~Adinolfi$^{50}$,
C.A.~Aidala$^{77}$,
Z.~Ajaltouni$^{7}$,
S.~Akar$^{61}$,
P.~Albicocco$^{20}$,
J.~Albrecht$^{12}$,
F.~Alessio$^{44}$
% ...

{\footnotesize \it

% ...
$ ^{7}$Universit{\'e} Clermont Auvergne, CNRS/IN2P3, LPC, Clermont-Ferrand, France\\
% ...
$ ^{12}$Fakult{\"a}t Physik, Technische Universit{\"a}t Dortmund, Dortmund, Germany\\
% ...
$ ^{20}$INFN Laboratori Nazionali di Frascati, Frascati, Italy\\
% ...
$ ^{29}$Nikhef National Institute for Subatomic Physics, Amsterdam, Netherlands\\
% ...
$ ^{43}$Instituto Galego de F{\'\i}sica de Altas Enerx{\'\i}as (IGFAE), Universidade de Santiago de Compostela, Santiago de Compostela, Spain\\
$ ^{44}$European Organization for Nuclear Research (CERN), Geneva, Switzerland\\
% ...
$ ^{46}$Physik-Institut, Universit{\"a}t Z{\"u}rich, Z{\"u}rich, Switzerland\\
% ...
$ ^{50}$H.H. Wills Physics Laboratory, University of Bristol, Bristol, United Kingdom\\
% ...
$ ^{61}$University of Cincinnati, Cincinnati, OH, United States\\
% ...
$ ^{63}$Syracuse University, Syracuse, NY, United States\\
% ...
$ ^{77}$University of Michigan, Ann Arbor, United States, associated to $^{63}$\\
% ...
}
\end{flushleft}
\end{document}